\documentclass[letter]{revtex4}
\usepackage[T1]{fontenc}
\usepackage{amsbsy}
\usepackage{amssymb}
\usepackage{graphicx}

\begin{document}

\title{Direct evidences for inner-shell electron-excitation by laser induced
electron recollision}

\author{Yunpei Deng$^{1,\dagger}$, Zhinan Zeng$^{2,3,\dagger}$ Zhengmao
Jia$^{2,\dagger}$, Pavel Komm$^{5}$, Yinhui Zheng$^{2}$, Xiaochun
Ge$^{2}$, Ruxin Li$^{2,3,4*}$ and Gilad Marcus$^{5,*}$}

\affiliation{$^{1}$SwissFEL, Paul Scherrer Institut, 5232 Villigen PSI, Switzerland.}

\affiliation{$^{2}$State Key Laboratory of High Field Laser Physics, Shanghai
Institute of Optics and Fine Mechanics, Chinese Academy of Sciences,
Shanghai 201800, China.}

\affiliation{$^{3}$IFSA Collaborative Innovation Center, Shanghai Jiao Tong University,
Shanghai 200240, China.}

\affiliation{$^{4}$School of Physical Science and Technology, ShanghaiTech University,
Shanghai 200031,China.}

\affiliation{$^{5}$Department of Applied Physics, The Benin School of Engineering
and Computer Science, The Hebrew University of Jerusalem, Jerusalem
91904, Israel.}

\maketitle
\noindent \textbf{Extreme ultraviolet (XUV) attosecond pulses, generated
by a process known as laser-induced electron recollision, are a key
ingredient for attosecond metrology, providing a tool to precisely
initiate and probe sub-femtosecond dynamics in the microcosms of atoms,
molecules and solids\citep{krausz2009attosecond}. However, with the
current technology, extending attosecond metrology to scrutinize the
dynamics of the inner-shell electrons is a challenge, that is because
of the lower efficiency in generating the required soft x-ray $\left(\mathbf{\hbar\omega>300\mathrm{eV}}\right)$
attosecond bursts and the lower absorption cross-sections in this
spectral range. A way around this problem is to use the recolliding
electron to directly initiate the desired inner-shell process, instead
of using the currently low flux x-ray attosecond sources.} \textbf{Such
an excitation process occurs in a sub-femtosecond timescale, and may
provide the necessary ``pump'' step in a pump-probe experiment\citep{kling2006control}.
Here we used a few cycle infrared $\boldsymbol{\mathbf{(\lambda_{0}\approx1800\mathrm{nm})}}$
source\citep{zhang2009tunable} and observed direct evidences for
inner-shell excitations through the laser-induced electron recollision
process. It is the first step toward time-resolved core-hole studies
in the keV energy range with sub-femtosecond time resolution. }

Most of light-matter interaction processes, within the limits of ``low''
frequency and a few-photon interaction, are well described by the
single-active-electron approximation. In this approximation the dynamic
of only one valence electron is considered, while the effect of the
rest is taken as an averaged masking charged cloud. On the other hand,
excitations of electron from deeper shells are usually accompanied
by multi-electron dynamics such as double excitation, the Auger decay,
Cooper minima, and the giant resonance, which can not be explained
by the single active electron approximation. Such excitations are
unstable and usually decay on a timescale ranging from few femtosecond
to few attosecond\citep{drescher2002timeresolved,penent2005multielectron}.
The decay may take place in a single step, but more often occurs as
a cascade of radiative and non-radiative channels. Spectroscopic data
may give some general information about the nature of such dynamics
but often fail to follow the exact details, for example, the line
widths of the cascade Auger decays reveal the total decay rate but
not the order of decaying channels and their individual decay rate.
To really follow such dynamics, one resorts to a time domain spectroscopy\citep{uphues2008ionchargestate,verhoef2011timeandenergyresolved,penent2005multielectron},
in which a first ``pulse'' initiates the process and a second ``pulse''
probes it. Since the relevant timescale for such dynamics spans from
attosecond to femtosecond and the relevant energy-scale spans from
$10^{2}-10^{5}\mathrm{eV}$, x-ray attosecond bursts may be the choice
to serve as the pump and the probe events. However, with the low photon
flux of current soft x-ray attosecond sources $\left(\hbar\omega>300\mathrm{eV}\right)$
and the low absorption cross-sections in this spectral range, it is
currently impossible to both pump and probe these processes with attosecond
x-ray pulses. To study processes involving valence electrons, an ultrashort
infrared pulse is often used to initiate the process, and a well synchronized
XUV attosecond pulse probes it\citep{krausz2009attosecond}. It is
difficult to extend this scheme to excite inner-shell processes because
of the large energy difference between inner-shell energies and the
infrared photon energy. Excitation of inner-shell dynamics by laser-induced
electron recollision might be the solution. Here we show direct evidences
for such excitations, as opposed to previous indirect evidences\citep{marcus2012subfemtosecond,shiner2011probing}.
Such an excitation process occurs in a sub-femtosecond timescale,
thus, provide the necessary ``pump'' step and might become key for
future ``pump-probe'' studies of inner-shell dynamics.

\begin{widetext}

\begin{figure}
\includegraphics[bb=0bp 100bp 612bp 580bp,clip,width=16cm]{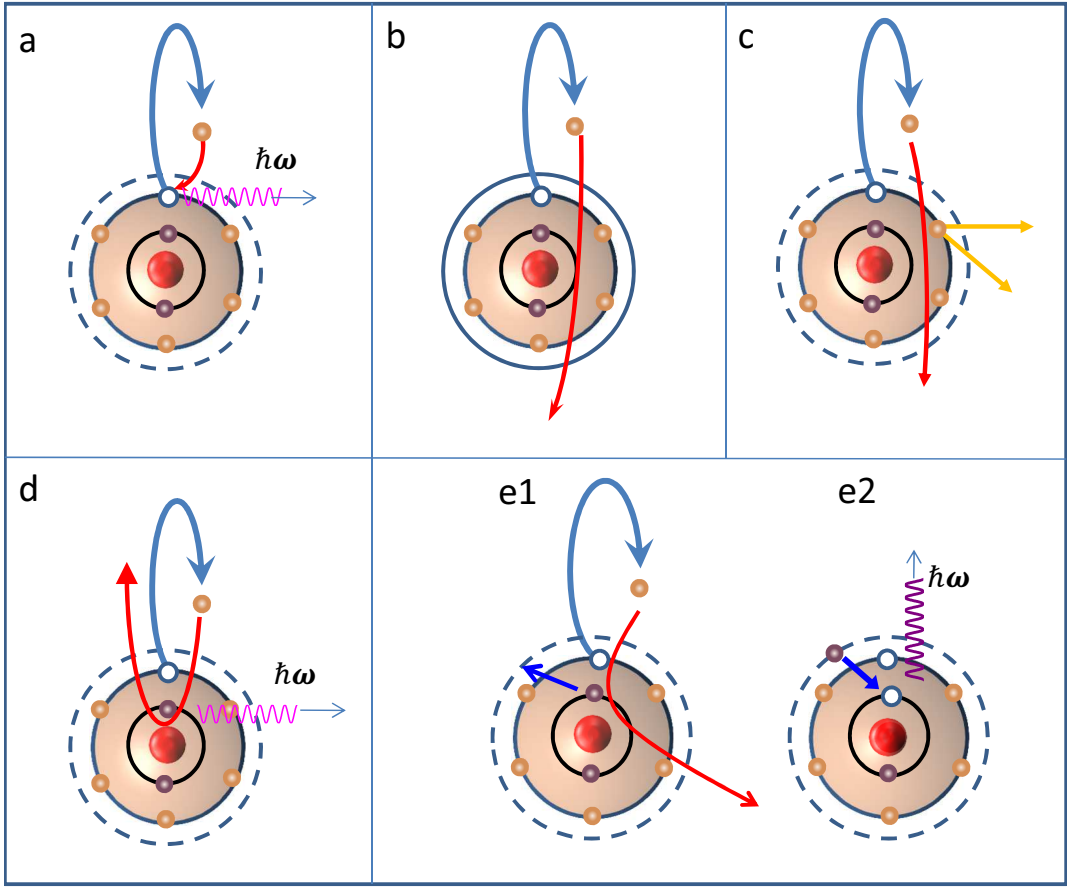}\protect\caption{\label{fig:Re-collision_channels} The different outcomes from the
laser-induced electron recollision process. a) the returning electron
recombines with the hole left in the valence shell and release its
excess energy as a XUV photon of the high order harmonics. b) the
electron is weakly scattered by the parent ion (elastic scattering)
leading to the discrete ATI spectrum. c) the returning electron kicks
out another electron (NSDI). d) A large-angle scattering of the electron
by the parent ion followed by soft x-ray emission due to the strong
charge acceleration (Bremsstrahlung radiation). e) the returning electron
excites or ionizes an inner shell electron, leaving a hole in this
shell (e1). After a while, this excited electron recombines with the
core-hole and may emit a soft x-ray photon or decay by another non-radiative
process (e2). }
\end{figure}

\end{widetext}

Insight to the laser-induced electron recollision process is given
in a semi-classical model\citep{krause1992highorder,corkum1993plasmaperspective},
in which, the electron is first tunnel-ionized by the strong electromagnetic
laser field and then it is accelerated forth and back by the same
alternating field to return to its parent ion with excess kinetic
energy. The outcome of this recollision may split into three different
channels (see figure \ref{fig:Re-collision_channels}): the first
is recombination with the parent ion while emitting energetic electromagnetic
radiation (high order harmonic and XUV attosecond pulse generation);
the second is elastic scattering, which manifests itself in a discrete
energy-spectrum of the scattered electron, known as above threshold
ionization (ATI \citep{eberly1991abovethreshold}); and the third
channel is an inelastic scattering, that results in additional excitation
or ionization. Indeed, only a few years after the discovery of the
ATI and the high order harmonic (HOH), the accompanying ``non-sequential
double ionization'' (NSDI) process was discovered\citep{lhuillier1983multiply,fittinghoff1992observation}.
While both the ATI and the HOH may well be explained in the single-active-electron
approximation through the above mentioned three-steps model, the NSDI
is the first evidence for laser induced inelastic recollision in which
the returning electron kicks out another electron. The same returning
electron may be used to initiate ultrafast processes in atoms and
molecules with a sub-femtosecond time resolution\citep{kling2006control}.
If the recolliding electron gains enough kinetic energy, it may also
initiate inner-shell excitations. 

\section*{Methods}

{In our experiment we focus a 12 fs, 1mJ infrared radiation
source $(\lambda_{0}=1.8\mathrm{\mathrm{\mu m})}$ on a pulsed gas-jet
and observed the soft x-ray radiation from the interaction region.
The reason for using infrared laser instead of 800 nm laser is the
quadratic scaling of the ponderomotive energy with the wavelength
of the drive laser. According to the semi-classical three steps model,
the maximum energy an electron may come back with, when it collides
with the parent ion, is equal to $3.17U_{p}\propto I_{0}\lambda^{2}$
.Here, $U_{p}$ is the ponderomotive energy , $I_{0}$ and $\lambda$
are the peak intensity and the central wavelength of the drive field
respectively. Because of this quadratic scaling , there is now a great
interest in developing and using few-cycle radiation sources having
longer wavelengths than 800nm. Such infrared sources have already
demonstrated the extension of HOH spectra towards the soft x-ray range
($\hbar\omega>1\mathrm{keV}$)\citep{marcus2012subfemtosecond}. The
quadratic dependent of the pondermotive energy with the drive wavelength
also proves to be useful in generating bright and compact incoherent
hard x-ray source, by focusing sub 100fs infrared pulses on a solid
target\citep{weisshaupt2014highbrightness}. The infrared radiation
source we have used here is based on an optical parametric amplifier,
described in detail elsewhere\citep{zhang2009tunable}. The soft x-ray
radiation that is coming from the excited atoms is a measure of the
amount of excitation, but in order to separate it from the accompanying
HOH radiation, we observed the soft x-ray radiation at a right-angle
to the infrared propagation direction. The infrared beam was focused
on the gas target placed inside a vacuum chamber by a f=300mm $\mathrm{CaF_{2}}$
lens to a spot size of about $65\mathrm{\mu m}$ FWHM, results in
a peak intensity of about $3\times10^{15}\mathrm{W/cm^{2}}$. The
corresponding ponderomotive energy is about 900eV, enough to excite
the K-shell of neon and the L-shell of krypton. For the gas target
we used jets of neon and krypton from a pulsed nozzle (series 9 Parker
nozzle, orifice diameter of $350\mu m$ and backing pressures ranging
from 1-10 bar). Soft x-ray spectra from the krypton and the neon atom
was recorded by a silicon drift detector (Amptek XR-100SDD) with $12.5\mathrm{\mathrm{\mu m}}$
beryllium window.}

\begin{figure}
\includegraphics[width=8cm]{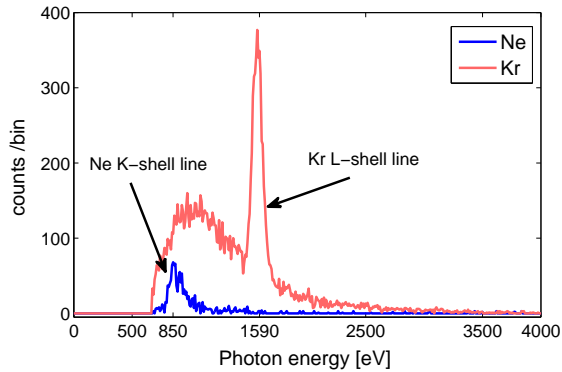}

\protect\caption{The fluorescence spectra from the excited Ne (Blue) and Kr (red) atoms.
The signal from the Ne is mainly from the K-shell transition with
a weak continuum that stretches up to $\approx1600\mathrm{eV}$. In
the Kr spectrum there is a sharp line, belonging to the L-shell transition
on top of a pronounced continuum. (the resolution of the SDD detector
is not high enough to resolve the $L_{\alpha},\,L_{\beta}$ splitting
).\label{fig:Ne-Kr_spectrum}}
\end{figure}

\section*{}

 Figure \ref{fig:Ne-Kr_spectrum} shows the soft x-ray spectra coming
from Ne and Kr targets. The spectrum shows the characteristic K-shell
line from Ne and L-shell line from Kr, on top of a wider continuum.
The continuum radiation from Kr extends up to \textasciitilde{}2800eV
which is in good agreement with the above mentioned formula for the
maximum possible kinetic energy of the re-colliding electron: $\mathrm{K}_{max}=3.17U_{p}=2.96\times10^{-13}I_{0}\lambda^{2}$
(K is in eV, $I$ is in $\mathrm{W/cm^{2}}$ and $\lambda$ is in
$\mu m$). We speculate that the origin of this continuum is either
coming from the recombination radiation or from the Bremsstrahlung
radiation. The sharp cutoff at the lower energy end is due to absorption
in the beryllium window.

Laser induced x-ray emission from solids and gas targets has a long
history, dating back almost to the dawn of lasers, when people start
to look at the interaction of nsec and psec lasers with plasmas. In
such a laser-matter interaction, the intense laser is heating the
plasma into a very high temperature through the inverse bremsstrahlung
(IB) or above-threshold-ionization (ATI) processes. The core-hole
excitations are in thermal equilibrium with the surrounding hot plasma.
X-ray emission is one of the channels to keep it in detailed balance.
For this model to be valid, it is required that the laser-induced
collision rate is much faster than the laser pulse duration, a condition
easily met with solid targets and picosecond long pulse duration.
As the pulse duration gets shorter and shorter and the target density
gets lower and lower, the conditions reach a certain point after which
there is not enough time for heating and thermalization and the resultant
plasma temperature drops to the level in which core-hole excitations
are not possible anymore. S. Dobosz et al. observed L-shell fluorecence
from Kr ions by using $5\times10^{17}\mathrm{W/cm^{2}}$, 130 fs,
$\lambda=800\mathrm{nm}$ laser, focused on pulsed gas-jet\citep{dobosz1997absolute}.
In electron impact excitation experiments it is known that the characteristic
emission may be un-isotropic if it results from transitions in which
the total angular momentum quantum number of the initial state is
j > 1/2. The angular distribution can then be expressed as $I\left(\theta\right)/I\left(\pi/2\right)=1-P\,cos^{2}\theta$
where \textit{P} is the polarization parameter\citep{mehlhorn1968onthe,kleinpoppen1980coherence}.
Bremsstrahlung radiation from projectile electrons may also show angular
distribution described by the modified Sommerfeld formula \citep{kulenkampff1959erganzungen,kleinpoppen1980coherence}:
$I\left(\theta\right)/I\left(\pi/2\right)=\left[\left(1-\beta cos\theta\right)^{2}-P\left(cos\theta-\beta\right)^{2}\right]/\left[\left(1-P\beta^{2}\right)\left(1-\beta cos\theta\right)^{4}\right]$.
In the case of excitations due to thermal equilibrium with a hot ambient
plasma, we would expect that this angular distribution will average
to zero. In their experiment, S. Dobosz et al. observed an isotropic
radiation pattern and therefore attributed this excitation to IB heating
of Kr clusters. C. Prigent et. al, investigated the dependence of
such excitations on the laser pulse duration\citep{prigent2008effectof}.
They found a sharp drop in the fluorescence yield as the laser pulses
get shorter than 50fs. 

\begin{figure}
\includegraphics[width=7cm]{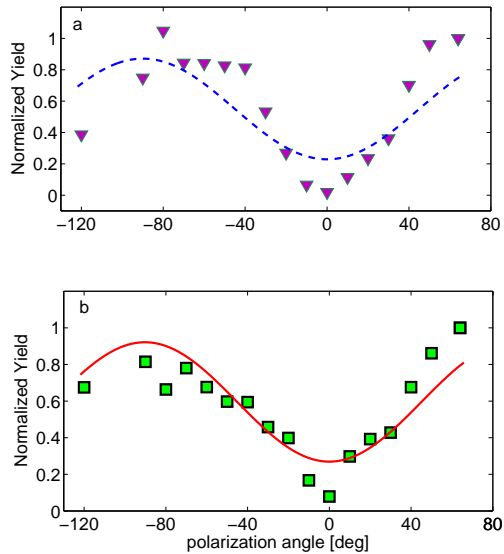}

\protect\caption{\label{fig:fluorescence-vs-angle}The krypton L-shell
fluorescence yield (a) and the continuum yield (b) as a function of
the angle between polarization direction and detector direction. We
fit the fluorescence yield to the $I\left(\theta\right)/I\left(\pi/2\right)=1-P\,cos^{2}\theta$
formula (blue broken line, P=0.716) and the continuum with the modified
Sommerfeld formula which is given above (red solid line, P=0.78)}
\end{figure}

The goal of this study is to show that the core-hole excitations we
have observed are indeed coming from the recollision process and not
from mere heating of the plasma. As a first step towards this goal
we followed the reasoning of \citep{dobosz1997absolute} and tested
the fluorescence and the continuum directionality. For that purpose
we used a $\lambda/2$ waveplate to rotate the polarization direction
with respect to the position of the detector and measured both the
x-ray fluorescence and the continuum yield as a function of that direction.
We observed a minimum in both of them when the polarized field pointed
toward the detector and a maximum when it was approximately perpendicular
to that direction (see figure\ref{fig:fluorescence-vs-angle}).
 With the pulse duration of our IR source (only 2 cycle) and the gas
densities we worked at ($10^{17}-10^{18}\,\mathrm{cm^{-3}}$ ), we
are not expecting IB and ATI heating to play an important role in
the observed core-hole excitations. The dipole like radiation pattern
(figure \ref{fig:fluorescence-vs-angle}) strongly support the recollision
excitation mechanism over the IB and ATI heating processes.

\begin{figure}
\includegraphics[width=7cm]{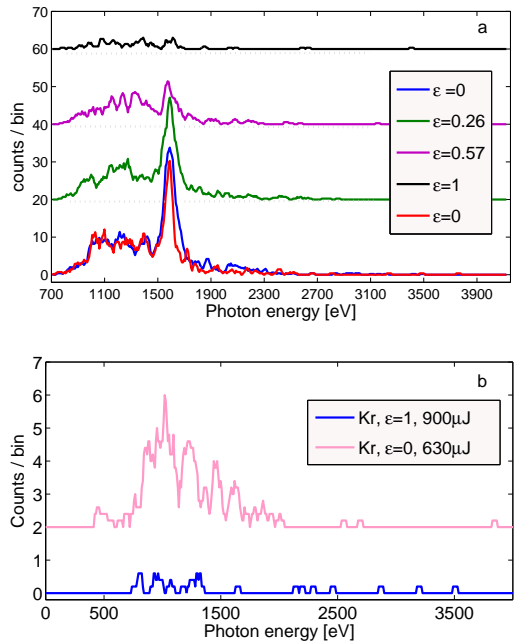} 

\protect\caption{a) The x-ray radiation yield from the Kr atoms vs. the drive ellipticity.
b) comparison of the x-ray yield when using a linearly polarized or
circlulary polarized infrared drive while keeping the peak electric
field the same. \label{fig:Yield-vs-ellipticity}}
\end{figure}

To further test whether we have recollision excitation or not, we
checked how the x-ray yield depends on the drive's ellipticity, since
the recollision process is highly sensitive to the polarization ellipticity
of the drive\citep{budil1993influence}. As the ellipticity gets larger
and larger, the electron trajectories are pushed away from the parent
ion and never come back to re-collide with it. Therefore, it is common
practice to check whether a process is coming from the recollision
process or not by changing the drive polarization ellipticity. figure
\ref{fig:Yield-vs-ellipticity}a shows the x-ray yield from the Kr
target as a function of the drive ellipticity.
 Indeed, we can see a strong reduction in the x-ray yield as the ellipticity
grows. However, as the ellipticity grows, the infrared peak intensity
is reduced and leads to a reduced ionization rate which can partially
explain the reduction of the photon yield. To check if the reduction
in the photon yield is due to the lower ionization yield or due to
the deflection of the returning electron, we compare again the results
from the linear polarization drive and the circular polarization drive,
this time we keep the peak electric field the same (see figure \ref{fig:Yield-vs-ellipticity}b).
This test shows clearly that the main reduction in the x-ray yield
is result of the electron deflection by the circular polarization.
Nevertheless, the signal is not completely disappear with circular
polarization as one would expect from the recollision induced excitation
model. The origin of this residual x-ray emission remain an open question,
calls for further investigation. This residual x-ray emission might
have a connection to other recently reported findings from other groups.
A non-sequential double ionization with circular polarization was
reported by a few groups\citep{gillen2001enhanced,guo2001ellipticity,bryan2006atomicexcitation,mauger2010recollisions}.
Mizuno et al. observed an extended tail in the spectrum of the photo-electrons
from the interaction of strong circular laser field with Kr atoms,
such a tail was believed to be a signature for electron recollision.
Another option is a''shake-up'' process, in which tunnel ionization
results in simultaneous excitation of one or more of the remaining
electrons\citep{bryan2006atomicexcitation}. It is also be possible
that the gas-jet is thick enough to allow for fast ATI electrons to
collide with atoms within the jet and excite them

In conclusion, the dipole-like radiation pattern (figure \ref{fig:fluorescence-vs-angle})
and the strong dependents of the x-ray emission on the drive ellipticity
(figure \ref{fig:Yield-vs-ellipticity}) are strong evidence supporting
the recollision excitation mechanism over the IB and ATI heating processes.
Since the recollision excitation process occurs at the sub-femtosecond
timescale, it opens the door for time domain studies of electron dynamics
in highly excited states where the recollision event initiates the
excitation followed by a synchronized attosecond probe pulse.

\section*{Author contributions}

G.M., Y.D., Z.Z. and R.L designed the experiment. G.M., Y.D., Z.J.,
P.K., Y.Z. and X.G. performed the experiment. G.M. and P.K. carried
out data analysis, R.L. provided experimental support and experiment
discussion. G.M., Y.D., Z.Z. wrote the manuscript, to which all authors
suggested improvment. G.M. supervised the project.
\begin{acknowledgments}
We would like to thank Ralph Ernstorfer for critical reading of the
manuscript and useful insights. GM acknowledge support by the Israel
Science Foundation (grant No. 404/12). ZZ and RL acknowledge support
by the National Natural Science Foundation of China (Grants No. 11127901,
No. 61221064 and No. 11222439) and National 973 project (No. 2011CB808103).
\end{acknowledgments}


\begin{thebibliography}{10}
	\expandafter\ifx\csname url\endcsname\relax
	\def\url#1{\texttt{#1}}\fi
	\expandafter\ifx\csname urlprefix\endcsname\relax\def\urlprefix{URL }\fi
	\providecommand{\bibinfo}[2]{#2}
	\providecommand{\eprint}[2][]{\url{#2}}
	
	\bibitem{krausz2009attosecond}
	\bibinfo{author}{Krausz, F.} \& \bibinfo{author}{Ivanov, M.}
	\newblock \bibinfo{title}{Attosecond physics}.
	\newblock \emph{\bibinfo{journal}{Rev. Mod. Phys.}}
	\textbf{\bibinfo{volume}{81}}, \bibinfo{pages}{163--72}
	(\bibinfo{year}{2009}).
	
	\bibitem{kling2006control}
	\bibinfo{author}{Kling, M.~F.} \emph{et~al.}
	\newblock \bibinfo{title}{Control of electron localization in molecular
		dissociation}.
	\newblock \emph{\bibinfo{journal}{Science}} \textbf{\bibinfo{volume}{312}},
	\bibinfo{pages}{246--248} (\bibinfo{year}{2006}).
	
	\bibitem{zhang2009tunable}
	\bibinfo{author}{Zhang, C.} \emph{et~al.}
	\newblock \bibinfo{title}{Tunable phase-stabilized infrared optical parametric
		amplifier for high-order harmonic generation}.
	\newblock \emph{\bibinfo{journal}{Opt. Lett.}} \textbf{\bibinfo{volume}{34}},
	\bibinfo{pages}{2730--2732} (\bibinfo{year}{2009}).
	
	\bibitem{drescher2002timeresolved}
	\bibinfo{author}{Drescher, M.} \emph{et~al.}
	\newblock \bibinfo{title}{Time-resolved atomic inner-shell spectroscopy}.
	\newblock \emph{\bibinfo{journal}{Nature}} \textbf{\bibinfo{volume}{419}},
	\bibinfo{pages}{803--807} (\bibinfo{year}{2002}).
	
	\bibitem{penent2005multielectron}
	\bibinfo{author}{Penent, F.} \emph{et~al.}
	\newblock \bibinfo{title}{Multielectron spectroscopy: The xenon 4d hole double
		auger decay}.
	\newblock \emph{\bibinfo{journal}{Phys. Rev. Lett.}}
	\textbf{\bibinfo{volume}{95}}, \bibinfo{pages}{083002}
	(\bibinfo{year}{2005}).
	
	\bibitem{uphues2008ionchargestate}
	\bibinfo{author}{Uphues, T.} \emph{et~al.}
	\newblock \bibinfo{title}{Ion-charge-state chronoscopy of cascaded atomic
		{Auger} decay}.
	\newblock \emph{\bibinfo{journal}{New Journal of Physics}}
	\textbf{\bibinfo{volume}{10}}, \bibinfo{pages}{025009}
	(\bibinfo{year}{2008}).
	
	\bibitem{verhoef2011timeandenergyresolved}
	\bibinfo{author}{Verhoef, A.~J.} \emph{et~al.}
	\newblock \bibinfo{title}{Time-and-energy-resolved measurement of {Auger}
		cascades following {Kr} 3d excitation by attosecond pulses}.
	\newblock \emph{\bibinfo{journal}{New J. Phys.}} \textbf{\bibinfo{volume}{13}},
	\bibinfo{pages}{113003} (\bibinfo{year}{2011}).
	
	\bibitem{marcus2012subfemtosecond}
	\bibinfo{author}{Marcus, G.} \emph{et~al.}
	\newblock \bibinfo{title}{Subfemtosecond {K}-shell excitation with a few-cycle
		infrared laser field}.
	\newblock \emph{\bibinfo{journal}{Phys. Rev. Lett.}}
	\textbf{\bibinfo{volume}{108}}, \bibinfo{pages}{023201}
	(\bibinfo{year}{2012}).
	
	\bibitem{shiner2011probing}
	\bibinfo{author}{Shiner, A.~D.} \emph{et~al.}
	\newblock \bibinfo{title}{Probing collective multi-electron dynamics in xenon
		with high-harmonic spectroscopy}.
	\newblock \emph{\bibinfo{journal}{Nat Phys}} \textbf{\bibinfo{volume}{7}},
	\bibinfo{pages}{464--467} (\bibinfo{year}{2011}).
	
	\bibitem{krause1992highorder}
	\bibinfo{author}{Krause, J.~L.}, \bibinfo{author}{Schafer, K.~J.} \&
	\bibinfo{author}{Kulander, K.~C.}
	\newblock \bibinfo{title}{High-order harmonic generation from atoms and ions in
		the high intensity regime}.
	\newblock \emph{\bibinfo{journal}{Phys. Rev. Lett.}}
	\textbf{\bibinfo{volume}{68}}, \bibinfo{pages}{3535--3538}
	(\bibinfo{year}{1992}).
	\newblock \bibinfo{note}{00957}.
	
	\bibitem{corkum1993plasmaperspective}
	\bibinfo{author}{Corkum, P.~B.}
	\newblock \bibinfo{title}{Plasma perspective on strong field multiphoton
		ionization}.
	\newblock \emph{\bibinfo{journal}{Phys. Rev. Lett.}}
	\textbf{\bibinfo{volume}{71}}, \bibinfo{pages}{1994} (\bibinfo{year}{1993}).
	
	\bibitem{eberly1991abovethreshold}
	\bibinfo{author}{Eberly, J.~H.}, \bibinfo{author}{Javanainen, J.} \&
	\bibinfo{author}{Rzazewski, K.}
	\newblock \bibinfo{title}{Above-threshold ionization}.
	\newblock \emph{\bibinfo{journal}{Physics Reports}}
	\textbf{\bibinfo{volume}{204}}, \bibinfo{pages}{331--383}
	(\bibinfo{year}{1991}).
	
	\bibitem{lhuillier1983multiply}
	\bibinfo{author}{L'Huillier, A.}, \bibinfo{author}{Lompre, L.~A.},
	\bibinfo{author}{Mainfray, G.} \& \bibinfo{author}{Manus, C.}
	\newblock \bibinfo{title}{Multiply charged ions induced by multiphoton
		absorption in rare gases at 0.53 mum}.
	\newblock \emph{\bibinfo{journal}{Phys. Rev. A}} \textbf{\bibinfo{volume}{27}},
	\bibinfo{pages}{2503} (\bibinfo{year}{1983}).
	
	\bibitem{fittinghoff1992observation}
	\bibinfo{author}{Fittinghoff, D.~N.}, \bibinfo{author}{Bolton, P.~R.},
	\bibinfo{author}{Chang, B.} \& \bibinfo{author}{Kulander, K.~C.}
	\newblock \bibinfo{title}{Observation of nonsequential double ionization of
		helium with optical tunneling}.
	\newblock \emph{\bibinfo{journal}{Phys. Rev. Lett.}}
	\textbf{\bibinfo{volume}{69}}, \bibinfo{pages}{2642} (\bibinfo{year}{1992}).
	
	\bibitem{weisshaupt2014highbrightness}
	\bibinfo{author}{Weisshaupt, J.} \emph{et~al.}
	\newblock \bibinfo{title}{High-brightness table-top hard {X}-ray source driven
		by sub-100-femtosecond mid-infrared pulses}.
	\newblock \emph{\bibinfo{journal}{Nat Photon}} \textbf{\bibinfo{volume}{8}},
	\bibinfo{pages}{927--930} (\bibinfo{year}{2014}).
	
	\bibitem{dobosz1997absolute}
	\bibinfo{author}{Dobosz, S.} \emph{et~al.}
	\newblock \bibinfo{title}{Absolute {keV} photon yields from ultrashort
		laser-field-induced hot nanoplasmas}.
	\newblock \emph{\bibinfo{journal}{Phys. Rev. A}} \textbf{\bibinfo{volume}{56}},
	\bibinfo{pages}{R2526--R2529} (\bibinfo{year}{1997}).
	
	\bibitem{mehlhorn1968onthe}
	\bibinfo{author}{Mehlhorn, W.}
	\newblock \bibinfo{title}{On the polarization of characteristic {X} radiation}.
	\newblock \emph{\bibinfo{journal}{Physics Letters A}}
	\textbf{\bibinfo{volume}{26}}, \bibinfo{pages}{166--167}
	(\bibinfo{year}{1968}).
	
	\bibitem{kleinpoppen1980coherence}
	\bibinfo{editor}{Kleinpoppen, H.} \& \bibinfo{editor}{Williams, J.~F.} (eds.)
	\emph{\bibinfo{title}{Coherence and {Correlation} in {Atomic} {Collisions}}}
	(\bibinfo{publisher}{Springer US}, \bibinfo{address}{Boston, MA},
	\bibinfo{year}{1980}).
	
	\bibitem{kulenkampff1959erganzungen}
	\bibinfo{author}{Kulenkampff, H.}, \bibinfo{author}{Scheer, M.} \&
	\bibinfo{author}{Zeitler, E.}
	\newblock \bibinfo{title}{Erg{\"a}nzungen zur {Sommerfeld}'schen {Theorie} der
		{R{\"o}ntgen}-{Bremsstrahlung}}.
	\newblock \emph{\bibinfo{journal}{Z. Physik}} \textbf{\bibinfo{volume}{157}},
	\bibinfo{pages}{275--281} (\bibinfo{year}{1959}).
	
	\bibitem{prigent2008effectof}
	\bibinfo{author}{Prigent, C.} \emph{et~al.}
	\newblock \bibinfo{title}{Effect of pulse duration on the x-ray emission from
		{Ar} clusters in intense laser fields}.
	\newblock \emph{\bibinfo{journal}{Phys. Rev. A}} \textbf{\bibinfo{volume}{78}},
	\bibinfo{pages}{053201} (\bibinfo{year}{2008}).
	
	\bibitem{budil1993influence}
	\bibinfo{author}{Budil, K.~S.}, \bibinfo{author}{Sali\`{e}res, P.},
	\bibinfo{author}{Perry, M.~D.} \&
	\bibinfo{author}{L{\textquoteright}Huillier, A.}
	\newblock \bibinfo{title}{Influence of ellipticity on harmonic generation}.
	\newblock \emph{\bibinfo{journal}{Phys. Rev. A}} \textbf{\bibinfo{volume}{48}}
	(\bibinfo{year}{1993}).
	
	\bibitem{gillen2001enhanced}
	\bibinfo{author}{Gillen, G.~D.}, \bibinfo{author}{Walker, M.~A.} \&
	\bibinfo{author}{Van~Woerkom, L.~D.}
	\newblock \bibinfo{title}{Enhanced double ionization with circularly polarized
		light}.
	\newblock \emph{\bibinfo{journal}{Phys. Rev. A}} \textbf{\bibinfo{volume}{64}},
	\bibinfo{pages}{043413} (\bibinfo{year}{2001}).
	\newblock \bibinfo{note}{00066}.
	
	\bibitem{guo2001ellipticity}
	\bibinfo{author}{Guo, C.} \& \bibinfo{author}{Gibson, G.~N.}
	\newblock \bibinfo{title}{Ellipticity effects on single and double ionization
		of diatomic molecules in strong laser fields}.
	\newblock \emph{\bibinfo{journal}{Phys. Rev. A}} \textbf{\bibinfo{volume}{63}},
	\bibinfo{pages}{040701} (\bibinfo{year}{2001}).
	\newblock \bibinfo{note}{00063}.
	
	\bibitem{bryan2006atomicexcitation}
	\bibinfo{author}{Bryan, W.~A.} \emph{et~al.}
	\newblock \bibinfo{title}{Atomic excitation during recollision-free ultrafast
		multi-electron tunnel ionization}.
	\newblock \emph{\bibinfo{journal}{Nat Phys}} \textbf{\bibinfo{volume}{2}},
	\bibinfo{pages}{379--383} (\bibinfo{year}{2006}).
	
	\bibitem{mauger2010recollisions}
	\bibinfo{author}{Mauger, F.}, \bibinfo{author}{Chandre, C.} \&
	\bibinfo{author}{Uzer, T.}
	\newblock \bibinfo{title}{Recollisions and correlated double ionization with
		circularly polarized light}.
	\newblock \emph{\bibinfo{journal}{Phys. Rev. Lett.}}
	\textbf{\bibinfo{volume}{105}}, \bibinfo{pages}{083002}
	(\bibinfo{year}{2010}).
	
\end{thebibliography}

\end{document}